# Randomized transmit and receive ultrasound tomography

G.T. Clement,[1,a)] and T. Kamakura[2]
[1]Department of Biomedical Engineering, Cleveland Clinic, Cleveland, Ohio USA
[2]The University of Electro-Communications, Chofu, Tokyo JAPAN

Abstract – A tomographic method is considered that forms images from sets of spatially randomized source signals and receiver sensitivities. The method is designed to allow image reconstruction for an extended number of transmitters and receivers in the presence noise and without plane wave approximation or otherwise approximation on the size or regularity of source and receiver functions. An overdetermined set of functions are formed from the Hadamard product between a Gaussian function and a uniformly distributed random number set. It is shown that this particular type of randomization tends to produce well-conditioned matrices whose pseudoinverses may be determined without implementing relaxation methods. When the inverted sets are applied to simulated first-order scattering from a Shepp-Logan phantom, successful image reconstructions are achieved for signal-to-noise ratios (SNR) as low as 1. Evaluation of the randomization approach is conducted by comparing condition numbers with other forms of signal randomization. Image quality resulting from tomographic reconstructions is then compared with an idealized synthetic aperture approach, which is subjected to a comparable SNR. By root-mean-square-difference comparisons it is concluded that - provided a sufficient level of oversampling - the dynamic transmit and dynamic receive approach produces superior images, particularly in the presence of low SNR.

a) gclement@physics.org

## I. INTRODUCTION

Having roots in x-ray crystallography, diffraction tomography was introduced in the early 1980s as a method for wave-based image construction.[1–4] Since then, numerous applications have been developed in the fields of optics,[5] electromagnetics,[6] and acoustics,[7] while improved methodology still continues to be an active area of research[8–11]. The primary advantages of the approach come from its potentially very high computational efficiency[12–14] as well as high spatial resolution[15]. Major limitations come in cases of highly variable acoustic wavenumbers[16] and multiple scattering, which generally must be treated via full wave inversion methods[17–26]. Such methods are considerably more computationally demanding[15], though it is noted that Wang et al[27] have recently reported on considerable gains in computing time using an encoding method.

In nearly all implementations, transmitters are treated as some arbitrary wave source (often a plane wave or point-source) that is shifted along some separable boundary, e.g. a ring or a line. Receivers are typically modeled as point-like and likewise positioned along a separable boundary[28]. The underlying assumption is that the sources (receivers) are capable of producing a constant and known beam (sensitivity) profile independent of position or rotation angle. In ultrasound, however, such consistency and knowledge is rarely achieved due to such factors as interelement coupling, differences in radiation patterns across transducer arrays[29] and scattering from outside the imaging region, including scattering from the transducer itself.

To address these limitations, an imaging technique is introduced which is devised around spatially-varied output signals and/or spatially varied receiver sensitivities. These variations may be fully controllable or may be a combination of controllable parameters (*e.g.* element amplitude and phase) and parameters inherent to the system (*e.g.* the abovementioned interelement coupling, etc.). The methodology considers transmitted and scattered signals on separable boundaries located external to the scattering object. Assuming a linear time-harmonic governing equation, it is understood that the overall contribution to the scattered signal at any point on a receiver boundary due to some point on a source boundary is a constant (unknown) factor, regardless of the degree of internal scattering. The received signal, in terms of this study, is understood to be the product between the (unknown) signal and the (controllable) spatially varying receiver sensitivity integrated over the over the receiver surface.

It will be shown that sets of spatially-varied source and sensitivity profiles can be produced that, if suitably selected, form invertible data sets that reduce the problem to that of idealized point-source/point-receiver pairs. In practice, such sets must be well-conditioned and must result in physically producible signals. Methods for generating sets through various types of randomization are considered where it is found that signals generated through the product of Gaussian and random functions consistently produce the desired behavior. The process is applied to tomographic simulations on ring and linear boundaries. It is found that, provided sufficiently high levels of oversampling, image reconstructions produce more accurate images, as compared to the idealized case - particularly in cases of low signal-to-noise ratio (SNR).

## II. THEORY

### A. Point sources and receivers

The relevant problem entails an arbitrary source that produces a time-harmonic signal $p_0(\mathbf{r}_S)$ at position $\mathbf{r}_S$, located on a separable boundary $\Gamma_S$ through which the entire incoming field passes. Beyond the boundary the field encounters an imaging region containing a spatially dependent scattering source $q(\mathbf{r})$ bounded by $Q$. Similarly, the entire outgoing scattered signal passes through some separable boundary $\Gamma_R$ with boundary value $p(\mathbf{r}_R)$. The two-dimensional cylindrical and Cartesian cases are illustrated in Fig. 1. Typically the transmitter is an array situated directly on $\Gamma_S$ producing a known incident field that is most often, but not necessarily[30], either planar or cylindrically spreading. Receivers are generally assumed to be point-like and situated directly on $\Gamma_R$ where scattering from $q(\mathbf{r}')$ is recorded and the process then repeated by translating (rotating) the source over the range of possible positions (angles) along $\Gamma_S$ to form the data $p(\mathbf{r}_R;\mathbf{r}_S)$ for the inverse problem. In the arguments to follow, the specific case of point sources and receivers under the Born approximation is relevant and stated here by[28]

$$p(\mathbf{r}_R;\mathbf{r}_S) = \int_Q p_0(\mathbf{r}';\mathbf{r}_S)q(\mathbf{r}')g(k_0,\mathbf{r}'|\mathbf{r}_R)\mathrm{d}^2\mathbf{r}'. \tag{1}$$

where $g$ is the free space Green's function of the background medium of wavenumber $k_0$ and $q(\mathbf{r}') = k_0^2 - k(\mathbf{r}')^2$.

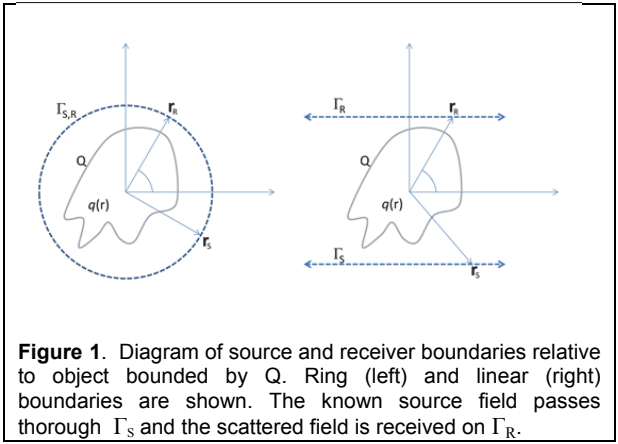

**Figure 1**. Diagram of source and receiver boundaries relative to object bounded by Q. Ring (left) and linear (right) boundaries are shown. The known source field passes thorough $\Gamma_S$ and the scattered field is received on $\Gamma_R$.

For a single point source at $\mathbf{r}_S$ and receiver at $\mathbf{r}_R$ it may be observed that, regardless of the complexity of the scattering, the signal reaching $\mathbf{r}_R$ must be proportional to the source, *e.g.* for the linear case doubling the source amplitude will double the receive amplitude or shifting the phase of the source will shift the received phase by an equal amount. This relation can be written in terms of a proportionality constant $T_{SR}$ between the source and receiver

$$p(\mathbf{r}_R;\mathbf{r}_S) = T_{SR} p(\mathbf{r}_S), \tag{2}$$

which will serve is a building block for the more general case below.

**B. The arbitrary case**

Considering now an arbitrary source over $\Gamma_S$, the source/receiver relation takes on the form of the summation of the point case (2),

$$p(\mathbf{r}_R;\mathbf{r}_S) = \sum_S^{\tilde{S}} T_{SR} p_0(\mathbf{r}_S), \tag{3}$$

summed over $\tilde{S}$ total points, and where the values $T_{SR}$ are generally unknown. The actual source $p_0(\mathbf{r}_S)$ might consist of an *inherent* component that is a function of geometry as well as imperfections such as spatial variation in surface velocity, load differences between elements, crosstalk, etc., as well as a *governable* component defined by controllable amplitude and phase over a given number of active elements. Though only governable components can be directly controlled, it will be assumed that the array is well characterized and both components are known. A third component, *noise*, will be defined as any remaining part of a measured signal.

Now, since $T_{SR}$ are independent of the incident field, any number of arbitrary fields may be generated under the identical proportionality terms, allowing some set of varied fields to be represented in matrix form as

$$p_{NR} = \hat{P}_{0_{NS}} T_{SR}, \tag{4}$$

where $T_{SR}$ takes the form of a column vector of length $\tilde{S}$, $\hat{P}_{0_{NS}}$ is a matrix containing $\tilde{N}$ rows of $p_0(\mathbf{r}_S)$ and $p_{NR}$ is an $\tilde{N}$-length column vector representing the scattered field values at $\mathbf{r}_R$. Extending (4) across all values on $\Gamma_R$, the equation takes the form of the matrix multiplication

$$\hat{P}_{NR} = \hat{P}_{0_{NS}} \hat{T}_{SR}. \tag{5}$$

Assuming $\tilde{N} \geq \tilde{S}$ and that $\hat{P}_{0_{NS}}$ is not singular, the values of $\hat{T}_{SR}$ can be determined by (pseudo) inversion,

$$\hat{T}_{SR} = \hat{P}_{0_{NS}}^{-1} \hat{P}_{NR}. \tag{6}$$

Comparing (6) and (2), it is recognized that for the particular set $\hat{P}_{0_{NS}}$ consisting of identical normalized signals independent of source location $\mathbf{r}_S$, then $T_{SR}$ and $p(\mathbf{r}_R;\mathbf{r}_S)$ must be identical. It is noted that, up to this point, no assumptions have been made about absorption or scattering within $q$, suggesting the relation holds even in cases of complex linear scattering. Under the Born approximation, however, it further follows that the solution of (6) is equivalent to the right hand side of (1) and therefore provides a formulation for evaluating reconstructions using varied source functions.

Next, the scenario is considered where a signal is received with variable sensitivity over some finite length of $\Gamma_R$ configured in a known, controllable pattern such that relative sensitivity at $\mathbf{r}_R$ is given by $A_{R\rho}$ with $\rho$ an index for a given pattern. As in the transmission case, this sensitivity will generally be a function of both inherent and governable components. A received pressure can be written as a summation over $\tilde{R}$ equivalent point receivers

$$p(\rho;\mathbf{r}_S) = \sum_{R}^{\tilde{R}} A_{R\rho} p(\mathbf{r}_R;\mathbf{r}_S) = T_{SR} p(\mathbf{r}_S), \tag{7}$$

where in contrast to the arbitrary transmission, $p(\mathbf{r}_R;\mathbf{r}_S)$ is now the unknown term. As with (4), the matrix form is developed over a set of $\tilde{\rho}$ sensitivity configurations. For a given point source S

$$p_{S\rho} = p_{SR} \hat{A}_{R\rho}, \tag{8}$$

where $p_{SR}$ is a row vector of length $\tilde{R}$ and where it will be demanded $\tilde{\rho} \geq \tilde{R}$. Over the set of $\tilde{S}$ point sources (8) expands to

$$\hat{P}_{S\rho} = \hat{P}_{SR} \hat{A}_{R\rho}. \tag{9}$$

As in the transmitted case, if the arbitrary receiver functions are appropriately selected to form a well-conditioned matrix, the values of $\hat{T}_{SR}$ can be determined by (pseudo) inversion,

$$\hat{P}_{SR} = \hat{P}_{S\rho} \hat{A}_{R\rho}^{-1}. \tag{10}$$

With cases entailing point receivers (6) and point transmitters (10) developed, it is straightforward to describe the more general case that considers both arbitrary transmitted fields and source sensitivity. The scattered fields (5) are now assumed to be recorded over receiver sensitivity profile $\hat{A}_{R\rho}$, then

$$\hat{P}_{N\rho} = \left[\hat{P}_{0_{NS}} \hat{T}_{SR}\right] \hat{A}_{R\rho} \quad (11)$$

Inverting the known matrices leads to

$$\hat{T}_{SR} = \hat{P}_{0_{NS}}^{-1} \left[\hat{P}_{N\rho} \hat{A}_{R\rho}^{-1}\right] \quad (12)$$

and (6) and (10) become special cases of (12).

In practice, the discretized functions forming $\hat{P}_{0_{NS}}$ and $\hat{A}_{R\rho}$ in (12) can be selected to form critically determined or overdetermined cases, which allow solutions to exist. Yet in all likelihood $\hat{P}_{0_{NS}}$ and $\hat{A}_{R\rho}$ will be ill-conditioned, indicating that, at best, approximate solutions might be obtained only after regularization procedures[31]. On the other hand, if source and receiver functions could be selected in some way to form well-conditioned matrices from the onset, reliable solutions to (12) might be obtained even in the presence of a noisy signal. Methodology therefore hinges on identifying a set of such functions that exist and that can also be generated in practice. If successfully identified, solutions to (12) could then be equated to their normalized point-like equivalents which, in turn, could be applied to established diffraction tomographic methods. An approach to generating these functions based on introducing a known but randomly selected component to the transmitted signal and receiver sensitivity is proposed and described in detail in section IIIB. Assuming, for now, that such solutions can be found, with the core approach of the study described by (12), application to the specific cases of cylindrical and linear boundaries can be considered.

### C. Cylindrical boundaries

For a point-like source, the unperturbed field in the integrand of (1) can be written in terms of the free space Green's function $p_0(\mathbf{r}';\mathbf{r}_S) = g(\mathbf{r}'|\mathbf{r}_S)$, which in cylindrical polar coordinates can be described in its separable form

$$g(\mathbf{r}'|\mathbf{r}_S) = \sum_n H_n(kr_S) J_n(kr') e^{in(\phi'-\phi_s)} \quad (13)$$

where $H_n$ represents the $n^{th}$ integer order Hankel function of the first kind and where $\{r',\phi'\}$ and $\{r_S,\phi_S\}$ are the cylindrical polar coordinates of $\mathbf{r}'$ and $\mathbf{r}_S$, respectively. Applying similar expansion to $g(\mathbf{r}'|\mathbf{r}_R)$, substitution into (1) gives

$$p(\mathbf{r}_R;\mathbf{r}_S) = \sum_{m,n} H_m(kr_S) H_n(kr_R) \int_Q J_m(kr') J_n(kr') e^{im(\phi'-\phi_S)} e^{in(\phi'-\phi_R)} q(\mathbf{r}') d^2\mathbf{r}' \quad (14)$$

When applied in Sec III, the left hand side of the equation is replaced by the equivalent $\hat{T}_{SR}$ calculated from (12). Rearranging and summing over the Fourier series with respect to $\phi_S$ and $\phi_R$ [32]

$$\sum_{m,n} \frac{P_{mn} e^{im\theta_S} e^{in\theta_R}}{H_m(kr_S) H_n(kr_R)} = \sum_{m,n} \int_Q J_m(kr') J_n(kr') e^{im(\phi'+\theta_S)} e^{in(\phi'+\theta_R)} q(\mathbf{r}') d^2\mathbf{r}'. \quad (15)$$

By the the Jacobi Anger expansion (15) becomes

$$\sum_{m,n} \frac{P_{mn} e^{im\theta_R} e^{in\theta_S}}{H_m(kr_S) H_n(kr_R)} = \int_Q e^{ikr'\sin(\phi'+\theta_R)} e^{ikr'\sin(\phi'+\theta_S)} q(\mathbf{r}') d^2\mathbf{r}'. \quad (16)$$

Substituting

$$x' = r'\cos\phi'$$
$$z' = r'\sin\phi' \tag{17}$$

into (16) the sin can be expanded as

$$r'\sin(\phi' + \theta_S) = x'\sin\theta_S + z'\cos\theta_S$$
$$r'\sin(\phi' + \theta_R) = x'\sin\theta_R + z'\sin\theta_R. \tag{18}$$

Further defining

$$k_x = -k(\sin\theta_S + \sin\theta_R)$$
$$k_z = -k(\cos\theta_S + \cos\theta_R) \tag{19}$$

then leads to the equation,

$$\sum_{m,n} \frac{P_{mn} e^{im\theta_S} e^{in\theta_R}}{H_m(kr_S)H_n(kr_R)} = \int_Q e^{-ik_x x'} e^{-ik_z z'} q(\mathbf{r}') dx' dz'. \tag{20}$$

so that by Fourier inversion

$$q(\mathbf{r}') = \int_Q \sum_{m,n} \frac{P_{mn} e^{im\theta_R} e^{in\theta_S}}{H_m(kr_S)H_n(kr_R)} e^{ik_x x'} e^{ik_z z'} dk_x dk_z. \tag{21}$$

**D. Linear boundaries**

The incident field can also be described over a line $\Gamma_S$ placed parallel to receivers along $\Gamma_R$, both lines being external to $q$. In Cartesian coordinates, an integral representation of the free space Green's function, the zeroth order Hankel function of the first kind, can be given by

$$g(\mathbf{r}|\mathbf{r}') = \frac{i}{4\pi} \int \frac{e^{ik_z|z-z'|}}{k_z} e^{ik_x'(x-x')} dk_x'. \tag{22}$$

where $k_z = \sqrt{k_0^2 - k_x'^2}$. Using this representation, (1) can be written as

$$p(\mathbf{r}_R;\mathbf{r}_S) = \int_Q \frac{e^{ik_{S_z}|z_S-z'|}}{k_{S_z}} e^{ik_{S_x}(x_S-x')} \frac{e^{ik_{R_z}|z_R-z'|}}{k_{R_z}} e^{ik_R'(x_R-x')} q(\mathbf{r}') dk_{S_x} dk_{R_x} d^2\mathbf{r}'. \tag{23}$$

for the case of a point source, where subscripts have been added to denote coordinates of S and R. By selecting the boundaries to be located along lines of constant z,

$$p(x_R;x_S) = \frac{i}{4\pi} \iint_{k_{S_x},k_{R_x}} Q(\mathbf{k}_R - \mathbf{k}_S) \frac{e^{ik_{R_z}z_R} e^{-ik_{R_z}z_S}}{k_{S_z}k_{R_z}} e^{ik_{R_x}x_R} e^{-ik_{S_x}x_S} dk_{S_x} dk_{R_x}, \tag{24}$$

where $Q$ is the Fourier transform of $q$ with respect to $x_S$ and $x_R$ leading to[28]

$$Q(\mathbf{k}_R - \mathbf{k}_S) = -2ik_{R_z} P(k_{R_x}, -k_{S_x}) e^{ik_{R_z}(z_S-z_R)} \tag{25}$$

which can finally be inverted to provide $q(\mathbf{r}')$.

## III. METHODS

### A. Synthetic data

Input for all calculations includes the wave number of the ambient medium and throughout the scattering region $q$, as well as the boundary geometry and associated grid sizes in the relevant coordinate system. For the cylindrical case, ring boundaries are parameterized in steps of equal angular spacing over scattering and receiver boundaries of constant radius. For linear boundaries equal linear steps are used, with the source and receiver lengths truncated at some specified distance on the boundaries. Free space Green's functions $G_S$ and $G_R$ are pre-calculated between boundary points and locations within the scatterer, which are then stored in a three-dimensional matrix format in RAM for efficient implementation of subsequent calculations. With the initial goal being to solve for the proportionality matrix $\hat{T}_{SR}$, a discretization of (1) is implemented to provide, for comparison, the ideal case representing a point source and point receiver.

In cylindrical coordinates, the Fourier series transform is introduced

$$p(r,\theta) = \sum_{m=-\infty}^{\infty} P_m(r) e^{im\theta} \tag{26}$$

such that the series coefficients are given by

$$P_m(r) = \frac{1}{2\pi} \int_0^{2\pi} p(r,\theta) e^{-im\theta} d\theta. \tag{27}$$

Simulation of an arbitrary source is performed by generating some $p_0(\mathbf{r}_S) = p_s(\mathbf{r}_S) + p_n(\mathbf{r}_S)$ on $\Gamma_S$, where $p_s$ is the total expected pressure inclusive of both inherent and governable quantities save some degree of random system noise $p_n$. Using (26) to expand $p_0(\mathbf{r}_S)$, the unperturbed field anywhere in $q$ may be expressed in terms of this series along with the expanded Green's function terms

$$p_0(\mathbf{r}'; \mathbf{r}_S) = \sum_m P_m(r_S) H_m(kr_S) J_m(kr') e^{im(\phi' - \phi_S)} \tag{28}$$

which can be substituted into the integrand of (1). In the Cartesian case, the Fourier transform of $p_0(\mathbf{r}_S)$ is given by

$$P(k_x; z_S) = \int_{-\infty}^{\infty} p(\mathbf{r}_S) e^{-ik_x x} dx \tag{29}$$

such that

$$p_0(\mathbf{r}'; \mathbf{r}_S) = \frac{1}{2\pi} \int_{-\infty}^{\infty} P(k_x; z_S) e^{i\sqrt{k^2 - k_x^2}(z - z_S)} e^{ik_x x} dk_x. \tag{30}$$

The scattered signal is simulated by substituting (28), (30), and $q$ into (1) and the signal is calculated on $\Gamma_R$ at half-wavelength increments. A sensitivity pattern $A(\mathbf{r}_R)$ is generated, representing both fixed characteristics of the receiver and controllable parameters, allowing the "received" signal to be calculated by summing the product of the scattered signal and the spatially-dependent sensitivity over $\Gamma_R$. This process is repeated using a series of $\tilde{N}$ varied source functions and $\tilde{\rho}$ varied sensitivity functions to form, by (11), the dataset representing $\hat{P}'_{N\rho} = \hat{P}_{N\rho} + p_{n_{N\rho}}$, in which additional receiver noise $p_{n_{N\rho}}$ has been added.

### B. Algorithm

#### 1. Source and receiver functions

The foremost task of the algorithm is to provide a reasonable estimate of solution $\hat{T}_{SR}$ in (12). As noted above in Sec. IIB, successful implementation is highly dependent upon selecting and generating source and receiver functions that can form well-conditioned $\hat{P}_{0_{NS}}$ and $\hat{A}_{R\rho}$. It has been established that randomized matrices are generally associated with low condition numbers[33]; a property that has been utilized for developing methods of solving linear problems[34]. Moreover, prior theoretical and experimental observations have indicated randomization can provide invertible solution sets for ultrasound applications[35,36] through temporal frequency randomization. Here, an approach for directly producing well-conditioned time-harmonic data is proposed through the use of the Hadamard (entrywise) product[37]. $S \circ T|_{i,j} \equiv S_{i,j} T_{i,j}$, between some preselected series of Gaussian-shaped discretized functions representing $p'_S$ or $A'_R$, and a random matrix $\Re$ of the same dimensions, whose elements consist of uniformly distributed pseudorandom numbers:

$$p_S = p'_S \circ \Re$$
$$A_R = A'_R \circ \Re. \qquad (31)$$

Each matrix is evaluated for two criteria: (i) Whether the function is physically producible, as determined by the constraints of a given source and receiver, and (ii) whether the matrix is well-conditioned as determined by condition number[37]

$$\kappa_A = \|A_R\| \|A_R^{-1}\|$$
$$\kappa_S = \|p_S\| \|p_S^{-1}\| \qquad (32)$$

under the criterion that the number be less than some pre-selected value. Brackets in (32) indicate matrix normal. Provided these criteria are met, synthetic data $\hat{P}'_{N\rho}$ are generated as described in Sec. IIIA. Pseudoinverses of the known matrices $\hat{P}_{0_{NS}}$ and $\hat{A}_{R\rho}$ are calculated via the Moore-Penrose Method[38] and, along with $\hat{P}'_{N\rho}$ as input, these values are used to solve for $\hat{T}_{SR}$ in (12).

#### 2. Image construction

For circular boundaries, (26), the Fourier series transform of $\hat{T}_{SR}$ is assumed equal to $P_{mn}$ in (21) which can then be directly solved to provide the image $q'(\mathbf{r}')$. For Cartesian boundaries reconstruction via (25) requires a nonlinear mapping between $(k_{R_x}, k_{S_x})$ and the Cartesian plane $(k_x, k_z)$ over which $\mathcal{Q}'(\mathbf{k}_R - \mathbf{k}_S)$ is specified. As previously detailed[14], this can be efficiently performed by solving for $k_{R_x}$ and $k_{S_x}$ in terms of $k_x$ and $k_z$ which have the solutions

$$k_{R_x} = \frac{k_x}{2} \pm \frac{1}{2}\sqrt{\frac{4k^2 k_z^2 - k_x^2 k_z^2 - k_z^4}{k_x^2 + k_z^2}}$$
$$k_{S_x} = k_{R_x} - k_x,$$
(33)

where if $k_{R_x} > k_x$ ($k_{S_x} > 0$) the additive solution (+) is used for Quadrants II and IV, the subtractive solution (-) for quadrant III, and no solution exists for quadrant IV. When $k_{R_x} < k_x$ ($k_{S_x} < 0$) the subtractive solution is used in quadrants I and III, the additive solution in quadrant IV, and no solution exists in quadrant II. Mapped data are then inverse-transformed to provide $q'(\mathbf{r}')$.

### C. Tests

#### 1. *Source and receiver functions*

To examine the effectiveness of the proposed randomization method given by (31), the approach is compared with alternative forms of signal randomization including: (i) random variation of amplitude; (ii) random variation of FWHM; (iii) randomized signal location. The terms "amplitude type," "FWHM type," "location type" and "Hadamard type" of randomization are introduced for identification of (i)-(iii) and (31), respectively. All permutations of the four types (16 combination types) are studied by generating 100 sets of each combination. Condition numbers are calculated using (32) and the mean and standard deviation for each type is determined. All tests are performed on data sets of length 1000 X 500. Two non-randomized FWHM values (5 and 10 row points) are examined in separate studies.

#### 2. *Image construction*

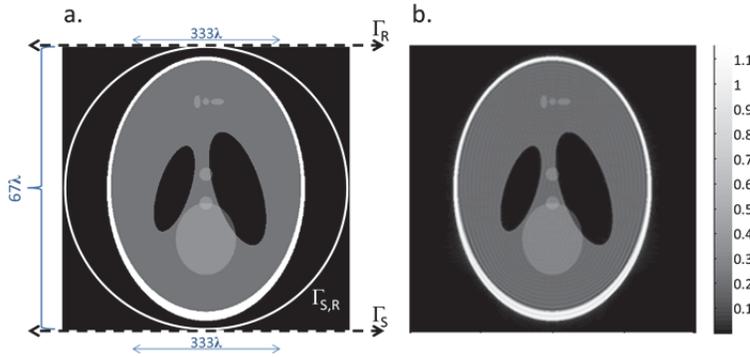

**Figure 2 (a)** Phantom dimensions relative to the linear and ring boundaries studied. (b) Reconstruction from the ring boundary under near-noiseless conditions (SNR>1000) constructed ring boundary data, and a FWHM of 10λ.

A Shepp-Logan numeric phantom[39] (Fig. 2a) is used to test the approach for both linear and ring-shaped boundaries in order to evaluate tomographic reconstructions as a function of SNR under varied input parameters. Source functions are generated by simulating an array positioned directly on a source boundary consisting of uniformly-radiating elements of given width $w$. Data are formed by selecting an element position about which a Gaussian-shaded pattern of a given amplitude and FWHM is centered. A set of $\tilde{N}$ such patterns is similarly generated to form the matrix set $\hat{P}_{0_{NS}}$ which is then augmented by a matrix of uniformly distributed pseudorandom values $\Re \in \{0,1\}$ using (31). Amplitude variation due to element-specific response is simulated by assigning a constant randomly-selected fractional response value to each element, formed as a column vector $\alpha_S$ that is entrywise multiplied to rows of $\hat{P}_{0_{NS}}$. The identical methodology is used to generate $\hat{A}_{R\rho}$. Once generated, $\hat{P}_{0_{NS}}$ and $\hat{A}_{R\rho}$ evaluated to determine whether the matrix condition numbers $\kappa_S$ and $\kappa_A$ are below a prescribed cutoff. Defining the overall SNR of the "measured" signal $\hat{P}'_{N\rho}$ as

$$\mathrm{SNR}_{N\rho} = \frac{\langle \hat{P}'_{N\rho} \rangle}{\sigma(p_n)}, \qquad (34)$$

with brackets indicating spatial mean and $\sigma$ the standard deviation, image reconstructions are evaluated as a function of SNR for a range of set sizes N and FWHM. For comparison, noise is similarly added to the idealized cases of point sources and point receivers, and SNR is determined by the "measured" set $\hat{P}'_{RS}$ as

$$\mathrm{SNR}_{RS} = \frac{\langle \hat{P}'_{RS} \rangle}{\sigma(p_n)}. \qquad (35)$$

For each simulation noise levels are adjusted to assure an equivalent SNR between (34) and (35). Root-mean-square deviations for an $\tilde{M}_x \times \tilde{M}_z$ image

$$\mathrm{RMSD} = \sqrt{\frac{\sum_{m_x,m_z}^{\tilde{M}_x \tilde{N}_z}(q_{m_x m_z} - q'_{m_x m_z})^2}{\tilde{M}_x \tilde{M}_z}} \qquad (36)$$

are calculated for all tomographic reconstructions.

Simulations are performed in scaled units of the wavelength in the ambient medium, $\lambda$. A 512X512 phantom spanning 67$\lambda$ in both directions with a spacing of 0.13$\lambda$. For the ring boundary, points on $\Gamma_S$ and $\Gamma_R$ are selected to be concyclic, with radius $a$ = 33$\lambda$ centered about the phantom leading to an angular spacing of $\Delta\phi_S = \Delta\phi_R = 0.015$ rad for both boundaries. An element size of 0.41$\lambda$ is considered, leading to 512 source/receiver elements over the boundary. For the linear boundaries, source and receiver lines are centered about phantom's z-axis extending a total length of 333$\lambda$ as shown in Fig. 2a. An element size of 0.5$\lambda$ is used resulting in 667 transmitters and receivers.

Green's functions $G_S$ and $G_R$ between source points and all points within the phantom, as well as between receivers and points in the phantom, are calculated and stored in two 512X512X512 three-dimensional matrices. Overdetermined set sizes are examined in even multiples of the number of source/receiver elements ($\tilde{R}=\tilde{S}$) ranging from $\tilde{N}=2\tilde{S}$ to $\tilde{N}=20\tilde{S}$ for the cases of FWHM = 10$\lambda$ and FWHM = 1.5$\lambda$. For each set, 512X512 images are formed under the presence of 11 different SNR levels, representing the integer range from 1 to 10 and one *very low* noise case (SNR > 1000). Prior examination of the matrices sizes in the study confirmed consistently well-conditioned matrices resulted when $\kappa \sim < 10^2$ so that a cutoff condition number 50 is selected for the tests.

## IV. RESULTS

Low condition numbers of the relevant randomized matrices were found to depend strongly upon the whether or not Hadamard type randomization was used. In fact, all matrices lacking the Hadamard type were found to be near singular. This behavior is evident in Fig 3a showing results for the 10-element FWHM case. For brevity, randomization combinations are denoted in the form R[h,a,l,w] indicating use of (h) Hadamard, (a) amplitude (l) Gaussian position, or (w) FWHM as 1, or as 0 if not used.

Lowest overall condition numbers were found to occur when the Hadamard type was used without other combinations (Fig 3b). Mean values of all combinations were: R[1111] = 232±20, R[1110] = 213±19, R[1101]=87±25, R[1100]=51±6, R[1011]= 144±9, R[1010] = 133±8, R[1001] = 44±9 R[1000] = 26±2, where uncertainty values are equal to ± one standard deviation of the data. Similar behaviors were observed in the 5-point FWHM case, where again absence of the Hadamard type resulted in near singular matrices while for the Hadamard cases R[1111] = 250±33, R[1110] = 217±25, R[1101]=170±**, R[1100]= 47±8, R[1011]=

154±16, R[1010] = 133±11, R[1001] = 79±** R[1000] = 21±2. A notable exception was found in the standard deviations for cases R[1101] and R[1001], where repeated runs found these cases consistently prone to outlier singularities making standard deviations highly variable between sets. They are thus omitted here, as indicated by ** above.

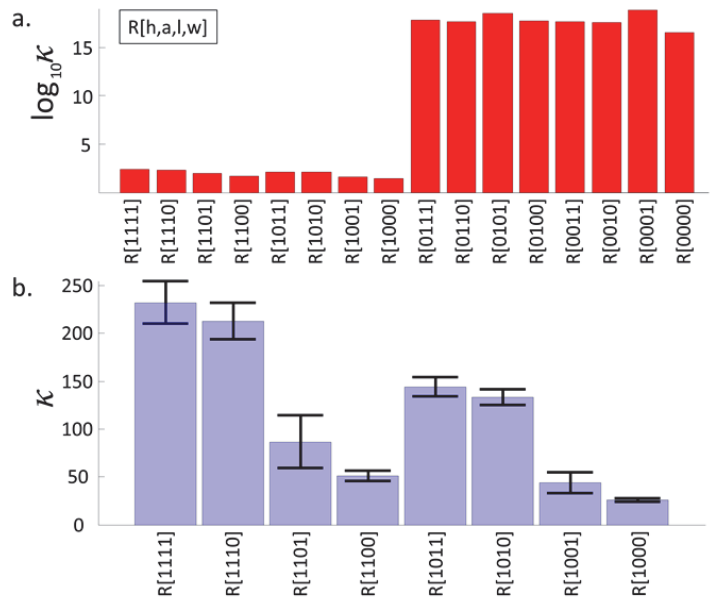

Figure 3. (a) Mean condition numbers are shown for matrix sets formed by 16 types of randomized parameters plotted on a log scale, Each bar is formed from 100 matrices of a given type. The first 8 types all use Hadamard types the last 8 types lack this randomization. (b) The first 8 types in (a) plotted on a linear scale. Errorbars represent ± 1 standard deviation. Types: h = Hadamard, a = Amplitude, l = position, w = FWHM.

Based on the results, images were subsequently produced using only Hadamard type of randomization. Examples of the actual element patterns produced by the approach are provided in Fig. 4 showing the $123^{rd}$, $133^{rd}$ and $143^{rd}$ randomized values taken from a set generated for the linear boundary case when FWHM = 1.5λ. Analysis of images derived from ring boundary data indicated RMSD values that decreased with increasing SNR (Fig. 5). In low SNR cases, results were found to depend strongly dependent on set size $\tilde{N}$, the larger $\tilde{N}$ having lower RMSD, and accordingly better image quality. For example, with SNR = 1, visual inspection found no discernible image for the $\tilde{N}=2\tilde{S}$ case, yet at $\tilde{N}=20\tilde{S}$ images were formed at all SNRs. Further inspection of $\tilde{N}=20\tilde{S}$ reconstructions revealed noisy images indiscernible from the very high SNR case for all SNR above 2. For both the randomized sets and the reference case (point-source/point-receiver), a lower limit of RMSD = 0.067 was observed at very high SNR levels. Values for data formed around a Gaussian FWHM = 10λ (Fig. 5a) were consistently higher than their equivalents formed around a FWHM = 1.5λ (Fig. 5b).

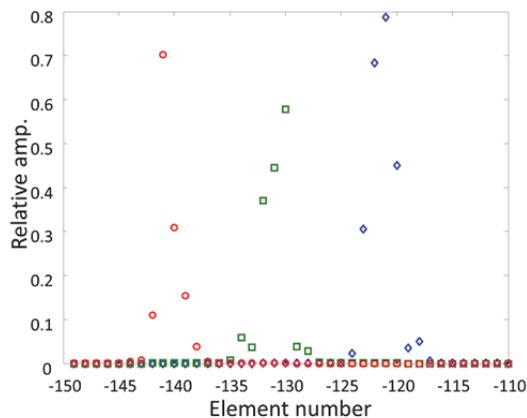

**Figure 4:** An example of Hadamard form randomization showing the $123^{rd}$, $133^{rd}$ and $143^{rd}$ randomized values from a set. Values are extracted from data used to construct the linear boundary case, FWHM = 1.5 λ. Element numbers are relative to a center element numbered 0.

Images from the linear boundary data revealed a systematic distortion in the line boundary data, such that even in the absence of noise images were distorted from actual values. With the FWHM set to 10λ a "best case" RMSD value was calculated to be 0.386, and all sample sizes were observed to trend negatively toward this limit as a function of SNR (Fig. 6). Once again, relative improvement was found when the FWHM was set to 1.5λ. In fact, in all randomized cases at this FWHM images were found to be superior to than those of the reference images. Moreover, for $\tilde{N}=8\tilde{S}$ and greater RMSD values were all lower than 0.392 for all SNR values studied.

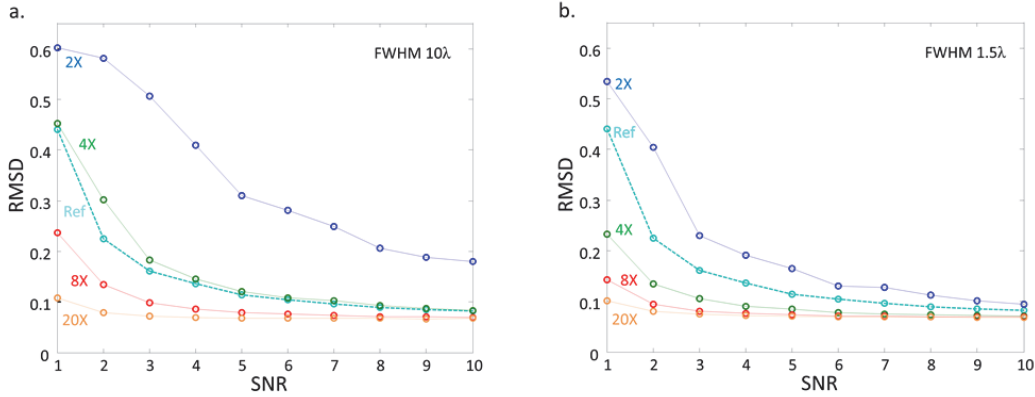

**Figure 5**. Root-mean-square deviation (RMSD) as a function of signal to noise ratio (SNR) for images formed using the Hadamard type of randomization of initially Gaussian-shaded sources and receivers. set at (a) FWHM = 10λ over a ring boundary and (b) FWHM = 1.5λ over the same boundary. Different curves represent datasets overdetermined by $2\tilde{S}$, $4\tilde{S}$, $8\tilde{S}$, and $20\tilde{S}$ along with a reference curve (Ref) created by idealized point and receiver sources.

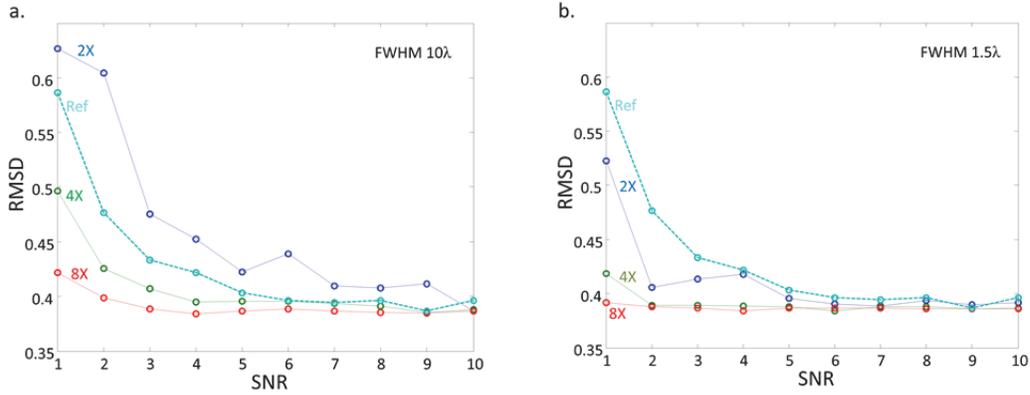

**Figure 6.** Root-mean-square deviation (RMSD) as a function of signal to noise ratio (SNR) for images formed using the Hadamard type randomization of initially Gaussian-shaded sources and receivers set at (a) FWHM 10λ and (b) FWHM 1.5λ over line boundaries. Curves represent datasets overdetermined by $2\tilde{S}$, $4\tilde{S}$, and $8\tilde{S}$ along with a reference curve (Ref) created by idealized point-sources and point-receivers.

## V. DISCUSSION

In the absence of Hadamard type randomization, all types of signal and receiver variation studied were found to produce near-singular data (Fig. 3). It was verified prior to testing that, as predicted, all cases lacking this randomization were unable to form images if any appreciable level noise was added to the simulated data. Conversely, in all randomization scenarios where the Hadamard type randomization was used, well-conditioned matrices formed in nearly all cases, thereby suggesting a signal approach that could make such the proposed multi-element method implementable in practice. Numeric tests of randomized datasets ultimately revealed a strong advantage to the use of Hadamard type alone without other forms randomization, and this type was consequently selected for image testing.

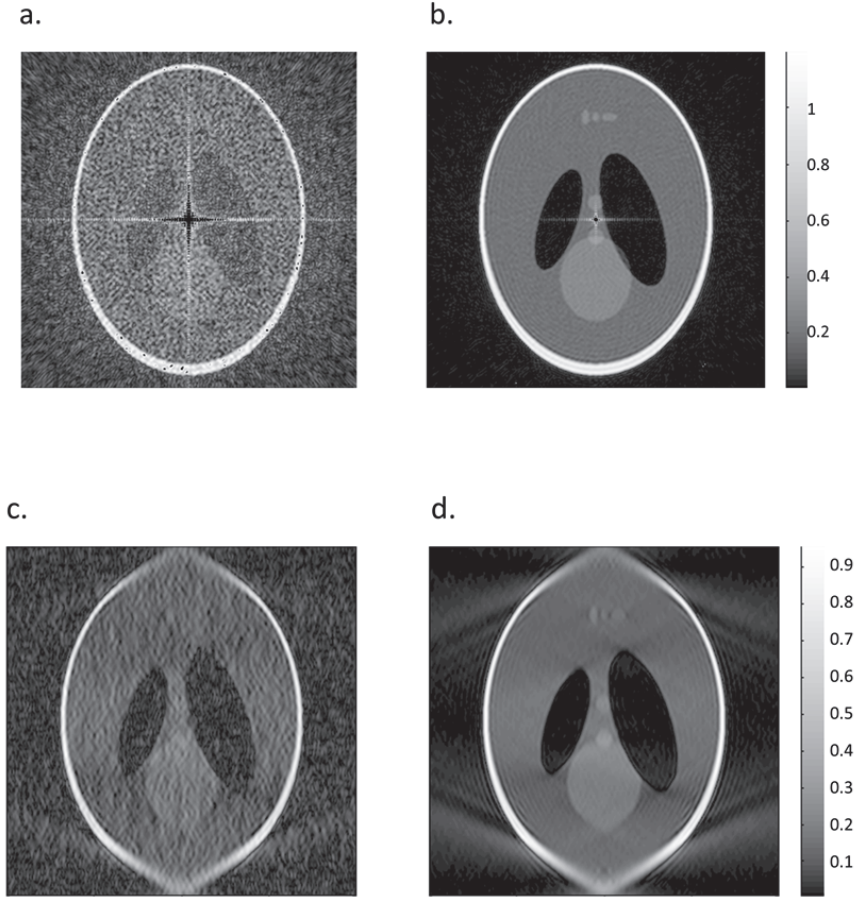

**Figure 7.** Image examples constructed in the presence of raw signals of SNR = 3, including: (a) an image formed from a point-source and point receiver on a ring boundary, (b) an image formed from randomized signals and receivers on a ring boundary (Hadamard form, $\tilde{N}=20\tilde{S}$, FWHM = 1.5λ) , (c) an image formed from a point-source and point receiver on opposing line boundaries, and (d) an image formed from randomized signals and receivers on opposing line boundaries (Hadamard form, $\tilde{N}=8\tilde{S}$, FWHM = 1.5λ) .

In imaging cases of low SNR, image quality was found to be highly dependent on the use of an increased number of signals $\tilde{N}$ , *i.e.* a higher degree of oversampling, to form images that were comparable to those of the very high SNR reference image. It was expected that the images would be comparable in RMSD to those of a simulated point-sources/point-receivers supplied with a comparable SNR and where the degree of oversampling was substituted by an equivalent number of signal averages. Interestingly, it was found that beyond the $\tilde{N}=4\tilde{S}$ oversampling case, the Hadamard type approach yielded lower RMSD values then the idealized case, particularly at the lowest SNR values studied. A qualitative examples of the resulting images of the point approach and the dynamic transmit and receive approach is provided when SNR=3 using the ring (Figs 7a,b) and linear (Figs 7c,d) configurations.

In addition to reduced RMSD, the method introduces additional potential imaging advantages. Through the use of multiple elements, increased gain is expected in both the source signal and overall receiver sensitivity. Signal strength alone could allow for considerable image improvement over point-like single element approaches. The arbitrary nature of the signals and receivers further provides ability to incorporate spatially-dependent factors such as variable element output, element coupling, and scattering from the array, as long as these factors are known *a priori*. Finally, although present testing of the approach was limited to first order scattering, the key methodologies given by (12) and (31) are valid for multiple scattering cases and might find application in full wave inversion.

## VI. SUMMARY AND CONCLUSIONS

This study introduced and numerically tested the concept of using random spatially varied signals to perform ultrasound tomography using an extended number of array elements for transmission and reception.

The method used multiple elements for transmission and reception which were reduced to an equivalent point-source/point-receiver case by inversion. This inversion process was found to be stable under noisy conditions in cases where a 'Hadamard type' randomization was used to manipulate the source signals and receiver sensitivity.

The algorithm may be summarized briefly as follows: 1) A set of driving signals and sensitivity profiles is selected by Hadamard type randomization, 2) sets are evaluated for condition number and producibility, 3) signals are generated and scattering recorded, 4) sets from 1) are numerically inverted and used with the signal data to solve equation (12) and 5) the solution is used in the relevant tomographic algorithm to form an image.

## VII.  ACKNOWLEDGMENT

This work was supported in part by The Center for Industrial and Governmental Relations, The University of Electro-Communications, and by award number R01EB014296 from the US National Institute of Biomedical Imaging and Bioengineering of the National Institutes of Health.